%
%
%
%
%
%
%
%
%
%
%
\magnification=1200
\font\t=cmcsc10 at 13 pt
\font\tt=cmcsc10
\font\n=cmcsc10
\font\foot=cmr9
\font\foots=cmsl9
\font\abs=cmr8
\font\babs=cmbx8

\centerline{\t On the bending of light from the Newtonian point of view\footnote{\dag}
{\foot{Memorie della Societ\`a Astronomica Italiana, (1923), 
{\bf 2}, pp.~371-379.}}}\bigskip
\centerline{Note by}\bigskip
\centerline{\n Attilio PALATINI}\bigskip
\vbox to 0.6 cm {}
\centerline{\tt translation and foreword by}\smallskip
\centerline{Marco Godina{\footnote{$^*$}{\foot SIGRAV, Viale F. Crispi 3 - 67100 L'Aquila (Italy).}}}\medskip
{\babs Foreword.} {\abs Within the framework of modified Newtonian gravity it is shown that the corrected Newtonian scheme can't explain both Mercury's perihelion advance and the bending of light at the same time.}

\vbox to 0.6 cm {}
It has been found by some authors \footnote{($^1$)}{\foot{Cf., e.g., F.~ANGELITTI,~}\foots{Sugli schemi newtoniani della gravitazione e sulla teoria della relativit\`a,}
\foot{Memorie della Societ\`a Astronomica, pp. 107-132 of the present Volume;~} \foot{L.~P.~EISENHART,~}\foots{The Einstein equations for the solar field from the Newtonian point of view~}\foot{(Science, 1922, vol.~LV, No.~1430, pp. 570-572);~}\foot{G. BERTRAND,~}\foots{La loi de Newton et la formule d'Einstein pour le p\' erih\' elie des plan\`etes.~}\foot{(Comptes Rendus, 1921, T.~173, pp. 438-440).}} how the secular shift in Mercury's perihelion can be explained by conveniently modifying the law of universal gravitation, without having to resort to the theory of relativity. In the new Newtonian scheme it should be assumed as expression of the unit central force, exerted by the mass $M$,
$$
F = -{fM\over r^2}\biggl(1+{3C^2\over{c^2r^2}}\biggr)\, ,\eqno(1)
$$
where $f$ is the constant of attraction, $c$ the speed of light and $C$ a new constant.
\par In proposing the law (1) is not said what value has to be assigned to $C$. But as, if the force is central, the integral of the areas exists, the constant $C$ is tacitly identified with twice the areal speed; which after all is required to the good ends that are to be achieved.
\par Mr.~Eisenhart in the note just mentioned above exactly, using the integral of areas, has bestowed to (1) the elegant form
$$
F = -{fM\over r^2}\biggl(1+3\,{{v_{\varphi}}^2\over{c^2}}\biggr)\, ,\eqno(2)
$$
where $v_{\varphi}$ identifies the transverse component of the speed of the moving point.
\par This author in regards to the law (2) says ``It would be interesting to know whether known discrepancies in the motion of the Moon would be overcome by the use of this law''.
\vfil
\eject
\par It might now be interesting to carry out a study in this sense; but, without going into complicated problems, we can deal with a more simple matter, related to the new universal gravitation law put forward.
\par One of the noteworthy phenomena, which pertains to the theory of relativity is notoriously that regarding the bending of light within a gravitational field. But it's certainly known, that a bending can be explained also regardless of Einstein's theory, by simply accepting the energy materialization principle, introduced in modern physics, irrespective of any special theoretical construction \footnote{($^1$)}{\foot{Cf. for the detailed development of this concept the following work, to which hereafter we will also have to refer to: T.   LEVI-CIVITA,~}\foots{Questions de mec\`anica cl\`assica i relativista~}\foot{(Institut d'Estudis Catalans, Barcelona: Conferencia IV, pg. 121 et  seq.),~}\foot{or~}\foots{L'ottica geometrica e la relativit\`a generale di Einstein~}\foot{(Rivista d'ottica e di meccanica di precisione, 1920) by the same author.~}}. If such principle is accepted then the light rays have to be considered as the trajectories of material particles and, as such, subject to the effect of a gravitational field.
\par If the universal gravitation law is Newton's, it is found that in the case of light rays, which skirt the edge of the sun, there must be a $0''.88$ bending, precisely half of the deviation required by the theory of relativity.
\par The direct observation (already carried out twice, as known) will say if the shift exists or not and to what extent.  Anyhow, we can questions ourselves: if the universal gravitation law is altered, always admitting that the energy is provided with weight, how is the $0''.88$ deviation correspondingly altered?
\par If by chance a $1''.75$ shift (the one foreseen by Einstein) had to be found, there would be the coincidence, of undeniable importance, of accounting, with a same law, for Mercury's perihelion anomaly and of the bending of light, without having to depend on the theory of relativity.
\par But such coincidence does not exist at all, as we will soon see. I thus believe that it's not the case of insisting on the proposal of modifying Newton's law, which, incidentally,  would be marred in its admirable simplicity, and that must stay, as it is, one of the most impressive edifices built in the field of natural philosophy \footnote{($^2$)}{\foot{As early as 1917, myself at pg. 48 of my Note~}\foots``{Lo spostamento del perielio di Mercurio e la deviazione dei raggi luminosi secondo Einstein''~}\foot{(Nuovo Cimento, 1917, Series VI, vol. XIV, pp. 12-54) I have given a form (not very different from (1)) which was useful to assign to the universal gravitation law in order to explain Mercury's perihelion anomaly. I did not want to modify Newton's law, but rather make use of a pure calculation artifice and show that, for the special problem considered, things went as if it were going an appropriate classical mechanics problem.}}.
\vfil
\eject
\par\medskip
{\tt 1.} Let's hence admit that the universal gravitation law shall be expressed  by (1). The force $F$ derives from the potential
$$
U = {fM\over r}\biggl(1+{C^2\over{c^2r^2}}\biggr)\, .
$$
\par In every mechanical problem, which depends from the potential $U$,  the integral of the areas will exist
$$
r^2{d\varphi\over dt}=rv_{\varphi}=C\, ,\eqno(3)
$$
and the energy integral
$$
{1\over 2}v^2-U=E\, .\eqno(4)
$$
\par For (3) the potential $U$ can also be written as
$$
U = {fM\over r}\biggl(1+{{v_{\varphi}}^2\over{c^2}}\biggr)\, .\eqno(5)
$$
\par\medskip
{\tt 2.} Let's now consider a  ray of light which propagates with velocity $c$ in a transparent medium, containing a force field deriving from the potential $U$.
\par If the point of view just now recalled concerning the materialization of energy is accepted, the rays of light have to be considered just as many   material particle trajectories. Each of which will hence have to behave as a free material point and obey, as such, to the laws of dynamics. In particular, due to the effect of the force deriving from the potential $U$, it will draw trajectories which, generally, will not be rectilinear.
\par But the laws that govern the movement of light, must not anyhow deviate much from the laws inferred from direct observation and that is: rectilinear propagation, at velocity $c$, within the gravitational field of the Sun and Earth. 
\par\medskip
{\tt 3.} Let's then admit that the gravitational field, within which a particle of light moves, shall be the one of the Sun, having mass $M$: its motion   shall obey to (3) and (4). 
\par Let's see which values shall have to be assigned to $E$ and $C$. 
\vfil
\eject
\par Let's start by assessing the order of magnitude of the potential $U$, provided by (5), where $v_{\varphi}$ has to be understood as the speed of the light's transverse component. 
\par The maximum value of the potential  ${fM\over r}$ is found when $r$ is identified with the Sun's ray $R$ and in this case we have \footnote{($^1$)}{\foot{Cf. T.~LEVI-CIVITA, l.~c., pg. 138. }}
$$
{1\over c^2}{fM\over R}=2.14\times 10^{-6}\, .\eqno(6)
$$
\par On the other hand it shall be noted that we do not know a priori how the light's velocity shall change within the Sun's field; but we do know that it must  differ only very slightly from $c$. Anyhow thus the factor  $1+{{v_{\varphi}}^2\over c^2}$  which appears in (5), or is smaller than $2$, or it differs from it  very slightly.  We can thus assume
$$
{1\over c^2}U = {fM\over c^2r}\biggl(1+{{v_{\varphi}}^2\over{c^2}}\biggr)=4.28\times 10^{-6}\, .
$$
\par The order of magnitude of ${1\over c^2}U$ is hence of a few millionths.
\par\medskip
Now in this degree of approximation,  (4) must lead to the first law of geometrical optics, i.e., propagation velocity equal  to $c$. So that this can take place it is sufficient to assign to the constant $E$ of the second member the value ${1\over 2}c^2$.
\par Indeed, in such a case, is had as exact expression of $v$,
$$
v^2 = c^2\biggl(1+{{2U}\over{c^2}}\biggr)\, ,
$$
i.e., save terms which are not at all negligible, 
$$
v = c\biggl(1+{{U}\over{c^2}}\biggr)\, .\eqno(7)
$$
\par And disregarding again ${{U}\over{c^2}}$ in respect to the unit
$$
v = c ,\quad Q.E.D.
$$
\vfil
\eject
\par\medskip
{\tt 4.} Let's now introduce the following hypothesis \footnote{($^1$)}{\foot{This hypothesis is not strictly necessary for the main purpose to which the present Note is aiming at. As it is undoubtedly  verified for the rectilinear motion it can always be asserted that at the perihelion it will be $v_{\varphi}= c\,(1 + \gamma)$ being $\gamma$ infinitesimal of the same order of magnitude as $\varepsilon$. Carrying out the calculations, is found that the equation (12) of the text, would be slightly modified, but not the general conclusion we could reach.}}:
\par When the light transits by a minimum distance point from the Sun, its speed is entirely transverse.
\par If we hence denote with $V$ the speed of light at the perihelion and with   $r_{o}$ such minimum distance, the (7) yields
$$
V = c\,\Biggl\{ {1+{fM\over {c^2r_{o}}}\biggl(1+{V^2\over{c^2}}\biggr)}\Biggr\}\, .\eqno(8)
$$
\par If one sets
$$
\varepsilon = {2fM\over{c^2r_{o}}}\, ,\eqno(9)
$$
$\varepsilon$ will in any case we are interested in be a very small quantity, of which we can neglect the square.
\par From (8) we will thus have
$$
V = c\,(1+\varepsilon)\, .
$$
\par Correspondingly from (3), applied to the perihelion, it will be
$$
r_{o}\,c\,(1+\varepsilon)=C\, .\eqno(10)
$$
\par\medskip
{\tt 5.} Let's now deal with the light trajectories' determination, which, in the absence of any perturbing circumstance, we know to be rectilinear.
\par The equation which yields the trajectories in a conservative problem concerning a potential $U$ and total energy $E$, is notoriously the following:
$$
\biggl({{du}\over {d\varphi}}\biggr)^2 + u^2 = 2\,{{{r_{o}}^2}\over {C^2}}\biggl(U + E\biggr)\, ,
$$
where $\varphi$ denotes the anomaly and $u ={{r_{o}}\over {r}}\,$.
\par In order to determine the shape of the light trajectories it will suffice to assume for $E$ the value ${1 \over 2}c^2$ and for $C$ the value given by (10).
\vfil
\eject
\par Hence we will have the equation
$$
\biggl({{du}\over {d\varphi}}\biggr)^2 + u^2 = {2\over {c^2(1+\varepsilon)^2}}\,\Biggl\{{{fM}\over {r_{o}}}u\biggl[1+u^2(1+\varepsilon)^2\biggr] + {1\over 2}c^2\Biggr\}\, ,
$$
that is, considering the position (9) and always neglecting the terms in $\varepsilon^2$
$$
\biggl({{du}\over {d\varphi}}\biggr)^2 = -u^2+\varepsilon\,(u+u^3)+1-2\varepsilon\, .\eqno(11)
$$
\par\medskip
{\tt 6.} Let's now integrate this equation to a first approximation.  
\par For $\varepsilon=0$ it is reduced to the form
$$
\biggl({{du}\over {d\varphi}}\biggr)^2 = 1-u^2\, ,
$$
whose integral is
$$
u = \cos\varphi\, ,
$$
choosing zero as the integration constant, as it's always legitimate when choosing the suitable polar axis (getting it to exactly coincide with the line joining the centre of the Sun with the perihelion of the light trajectories).
\par Let's now set
$$
u = \cos \varphi + \varepsilon \, \zeta
$$
and let's replace in (11) considering only the terms in $\varepsilon$. By this a linear equation in $\zeta$ is obtained:
$$
\sin \varphi \, {{d\zeta}\over {d\varphi}} = \zeta \, \cos \varphi -{1\over 2}\cos \varphi\, (1 + \cos^2 \varphi) + 1 \, ,
$$
which is immediately integrated and yields for $\zeta$ the value
$$
\zeta = 1 - \cos \varphi +{1\over 2}\sin^2 \varphi\, .
$$
\par The equations of the light trajectories, returning to $u$ its value ${{r_{o}}\over r}$, are hence provided by the equation
$$
{{r_{o}}\over r} = \cos \varphi + \varepsilon \, (1 - \varphi + {1\over 2}\sin^2 \varphi)\, .\eqno(12)
$$
\vfil
\eject
\par Such trajectories turn to be straight lines, normal to the polar axis,
$$
r_{o}  = r \cos \varphi \, ,\eqno(13)
$$
when (and only when) $\varepsilon = 0$, which amounts to saying when the effects of the Sun are neglected. But these differ from the straight lines (13) by a very small amount in the order of magnitude of $\varepsilon$.
\par Having written (12) in the form
$$
{{r_{o}}\over r} = \cos \varphi + 2 \, \varepsilon \sin^2 {\varphi \over 2} \, \biggl(1 + \cos^2 {\varphi \over 2}\biggr)\, ,
$$
it can be seen that the trajectories are symmetrical in respect to the polar axis, they are on the same side of the straight line (13) and turning their concavity towards the Sun.
\par\medskip
{\tt 7.} It's now easy to prove that the curves (12) admit two asymptotes (in  a symmetrical position in respect to the polar axis).
\par As our curves are given in polar coordinates, we can proceed in the following way. From the second member of the (12) it is found that the vector ray meets the curve at infinity at the two values of $\varphi$
$$
{\pi \over 2} + {{3\, \varepsilon} \over 2}\qquad , \qquad -{\pi \over 2} - {{3\, \varepsilon} \over 2}\, .
$$
\par If we now indicate with $\vartheta$ the angle formed by a generic tangent to the trajectory and by the vector ray passing through contact point, and with $\alpha$ the angle  formed  by this same tangent with the  polar axis, it is obtained, as is well known,
$$
\cot \vartheta = {1 \over r}\,{{dr} \over {d \varphi}}
$$
and thus also
$$
\cot (\alpha - \varphi) = {1 \over r}\,{{dr} \over {d \varphi}}\, .
$$
\par Always from (12) it is had
$$
\lim_{r \to \infty} {{1 \over r}\,{{dr} \over {d \varphi}}} = 0
$$
\vfil
\eject
\par\noindent and thus
$$
\cot \lim_{r \to \infty} (\alpha - \varphi) = 0\, .
$$
\par Follows
$$
\lim_{r \to \infty} \alpha = \cases{\;{\pi \over 2} + {{3\, \varepsilon} \over 2} \, ,\cr \, \cr \, \cr -{\pi \over 2} - {{3\, \varepsilon} \over 2} \, .\cr}
$$
\par Two asymptotes thus exist, being $\delta$ the acute angle which they form
$$
\delta = 3\, \varepsilon \, .
$$
\par\medskip
{\tt 8.} Let's now suppose that on earth the light coming from a star could be observed. The ray of light will bend according to a curve (12) and  thus  an apparent deviation of the star which the light originates from will have to be noticed. By which angle?
\par As both the star as well as the Earth can be considered at an infinitely great distance from the Sun, the bending will be measured by the asymptotes' angle $\delta$ just now determined.
\par Now, taking into account the expression of $\varepsilon$, and considering a ray which borders on the edge of the Sun (with this $r_{o}$ has to be identified with the radius $R$ of the Corona) is obtained
$$
\varepsilon = {2fM\over{c^2R}}\, ,
$$
that is for (6)
$$
\varepsilon=4.28\times 10^{-6}\, .
$$
\par It is thus
$$
\delta = 12.84\times 10^{-6}
$$
and reducing to arc seconds,
$$
\delta'' =12.84\times 10^{-6} \times {{6^4 \times 10^3} \over{2\, \pi}}  = 2''.64 \, ,
$$
that is {\it three times\/} the angular displacement required by the simple Newtonian law.
\par\medskip
{\tt 9.} The corrected Newtonian scheme can't therefore explain both Mercury's perihelion anomaly and the bending of light {\it at the same time\/}. If this phenomenon does exist (and of course to a different degree from

\vfil
\eject
\par
\noindent the one just now calculated) and the principle of materialization of energy is accepted, it has to admitted that the law of the attraction of light shall be  different from the law of attraction of the material masses as they are commonly called. In this way Einstein's wonderful synthesis of the mechanical and electromagnetic phenomena is lost and the problem of the search for the unique scheme arises again, that is the problem which gave rise to all the impressive physico-mathematical researches carried out during the last hundred years and that culminated in the creation of the theory of relativity.
\par\medskip Royal University of Parma, January 1923.

\vbox to 3.0 cm {}
\centerline{\tt --- --- --- ---}

\end